\documentclass[a4paper,11pt]{article}
\usepackage{aaskaiid}

\usepackage{orcidlink}

\usepackage{graphicx}
\usepackage{amsmath}
\usepackage{natbib}
\usepackage{aas_macros}

\title{Radio Diagnostics of Particle Acceleration in Solar Flares with SKAO Observations}
\ShortTitle{Observing particle acceleration in solar flare from SKAO}

\author[1]{Rohit Sharma \orcidlink{0000-0003-0485-7098}}
\ShortName{Sharma, R. et al.} 
\author[2]{Marina Battaglia \orcidlink{0000-0003-1438-9099}}
\author[3]{Sijie Yu \orcidlink{0000-0003-2872-2614}}
\author[4]{Hamish Reid \orcidlink{0000-0002-6287-3494}}
\author[5]{Eduard Kontar \orcidlink{0000-0002-8078-0902}}
\author[3]{Bin Chen \orcidlink{0000-0002-0660-3350}}

\affiliation[1]{Space, Planetary \& Astronomical Sciences \& Engineering (SPASE), Indian Institute of Technology Kanpur, Kalyanpur, Kanpur, 208016, Uttar Pradesh, India}
\emailAdd{rsharma@iitk.ac.in}
\affiliation[2]{Fachhochschule Nordwestschweiz, Bahnhofstrasse 6, 5210 Windisch, Switzerland}
\affiliation[3]{Center for Solar-Terrestrial Research, New Jersey Institute of Technology, 323 M L King Jr Blvd, Newark, NJ 07102-1982, USA}
\affiliation[4]{University College London, Mullard Space Science Laboratory, Holmbury Hill Rd, Dorking, 106, Holmbury St Mary, United Kingdom}
\affiliation[5]{School of Physics \& Astronomy, University of Glasgow, Glasgow G12 8QQ, UK}

\abstract{Particle acceleration is a fundamental astrophysical process occurring across diverse systems and scales, producing electromagnetic emission across all wavelengths. Radio bursts from astrophysical systems like active galaxy jets, solar flares, pulsars, etc., provide a probe into the emission mechanism and particle acceleration processes. Among all astrophysical phenomena, magnetically driven solar flares provide unique diagnostics of nonthermal particles due to the advantage of multiple spatial-temporal and spectral measurements. The subsequent emitted radiation spans various sections of the radio spectrum. Based on brightness temperature and spectrum, one can distinguish between bright plasma emission or 'nonthermal' emission. Nonthermal radio bursts in meter and microwave bands arise from suprathermal particles, while the surrounding plasma produces fainter thermal emission.

High spatial-temporal studies along with polarimetry will enable tracking of electron beams in evolving magnetic fields and constrain coronal properties such as magnetic strength, density, and temperature. Fine spectral structures from gyrosynchrotron bursts allow mapping of magnetic fields along acceleration tracks. However, emissions from the more numerous, weaker particle populations remain difficult to observe.

Particles in solar flares follow a power-law energy distribution, with weaker populations being more common. The Square Kilometre Array Observatory (SKAO) will provide high-fidelity data that will enable detailed characterisation of these populations, statistical studies of particle beams, and insights into their interaction with magnetic topologies. Its high-resolution (especially SKA-Mid) and multi-wavelength synergy (EUV, X-rays) will refine diagnostics of coronal and acceleration-region properties. This chapter reviews particle acceleration models and expected advances from SKAO.

}

\begin{document}
\maketitle

\section{Introduction}

The phenomenon of particle acceleration in plasma occurs throughout the universe in various astrophysical objects. It results in the production of emissions across the electromagnetic spectrum, including radio bursts, from sources such as active galaxies, solar flares, and pulsars, which serve as key probes of the underlying emission mechanisms and acceleration processes involved \citep[e.g.][]{2020NewAR..8901543M,2019Ap&SS.364..185U,2018ARNPS..68..377T,2015MNRAS.448..606C,2011SSRv..159..357Z,2024LRSP...21....1K}. Solar flares offer particularly valuable insights into nonthermal particles owing to extensive multi-wavelength, high-resolution spatial-temporal, and spectral observations \citep{2017LRSP...14....2B,2022Natur.606..674F,2011SSRv..159...19F}. The resulting radiation covers multiple bands of the electromagnetic spectrum. In the radio domain, direct diagnostics of particle acceleration regions during radio bursts can be obtained \citep{2008A&ARv..16....1P,2004P&SS...52.1399G,1985ARA&A..23..169D}. The specific parameters that can be constrained depend on the underlying emission mechanisms, predominantly plasma emission and gyro-synchrotron (GS) emission \citep{1986SoPh..104..207A, 2020FrASS...7...57N, 1979ApJ...234.1137D}. Previous studies have provided diagnoses of solar coronal plasma properties, including magnetic field strengths, electron densities, and plasma temperature, as well as providing insights into processes such as plasma heating and particle acceleration \citep{2005AdSpR..35.1759K,2011SSRv..159..225W,2022Natur.606..674F}.  In particular, detailed spectral profiles from gyrosynchrotron radio bursts will allow us to map magnetic fields along particle acceleration paths and build robust statistical datasets \citep{2021FrASS...7...77A}. With the high-fidelity imaging from the SKAO \citep{bourke2015advancing}. Therefore, we expect weaker emission of propagating particle beams can be detected by SKAO. 

Unlike other astronomical sources, high-resolution spatial and temporal observations of solar radio bursts have also made it possible to track electron beams within evolving magnetic field structures. Propagating particle beams may also be shaped by local magnetic topologies. Overall, combined with multi-wavelength data, including extreme ultraviolet and X-ray observations, SKAO results will significantly enhance diagnostics of the corona and particle acceleration sites. 

The paper is divided into six sections. The first section is the introduction (Sec. \ref{sec:pa}), followed by a description of particle acceleration in solar flares observed at radio wavelengths and the different types of associated radio events observed by radio spectrographs and interferometers. The third and fourth sections deal with ways to measure properties of particle acceleration and transport (Sec. \ref{sec:diagnostics}), and improvements expected from SKAO in a developmental phased manner (Sec. \ref{sec:ska}), respectively. The fifth section highlights the expected synergies with other solar observing instruments (Sec. \ref{sec:synergies}), followed by the summary in Section Six (Sec. \ref{sec:summary}).

\section{Particle Acceleration \& Radio emission}
\label{sec:pa}

\begin{figure*}[h]
    \centering
    \begin{tabular}{c}
	\includegraphics[width=0.4\columnwidth]{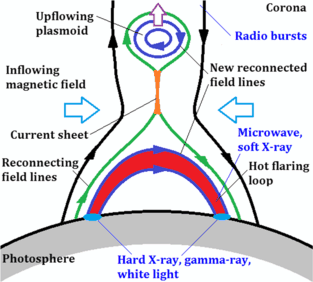} 
	\end{tabular}
    \caption{Standard model of the solar flare phenomenon. Credit: \cite{2016SSRv..200...75N} }
    \label{fig:sfm}
\end{figure*}

Solar flares are driven by the magnetic reconnection process, which can accelerate particles to high energies. These accelerated particles form the nonthermal particle population and cause intense release of radiation via various emission processes 
Magnetic field geometry present in the solar corona is mostly in the form of magnetic loops, which are anchored on the photosphere. In a basic scenario, the magnetic reconnection occurs at magnetically null points or quasi-separatrix layers, etc.  in the loop geometry \citep[e.g.][]{2006AdSpR..37.1269D}. These regions produce accelerated particles travelling towards (down) the chromosphere and away (up) from it. This picture of the solar flare phenomenon is called the `standard flare model' (Fig. \ref{fig:sfm}). However, intricate magnetic topologies involve complex particle acceleration and propagation during the solar flare \citep[e.g.][]{2017JPlPh..83a5301J}. The downward-travelling particles propagate along magnetic loops and emit gyrosynchrotron radio waves, and eventually hit the dense chromosphere, producing hard X-rays \citep{1985SoPh..100..465D}. The upward propagating particles can emit radiation, mostly seen from plasma emission in the form of radio bursts \citep{1982SSRv...32..405K,2023arXiv230707144K}. 

\subsection{Conventional Radio bursts}

\begin{figure*}[h]
    \centering
    \begin{tabular}{c}
	\includegraphics[width=1.0\columnwidth]{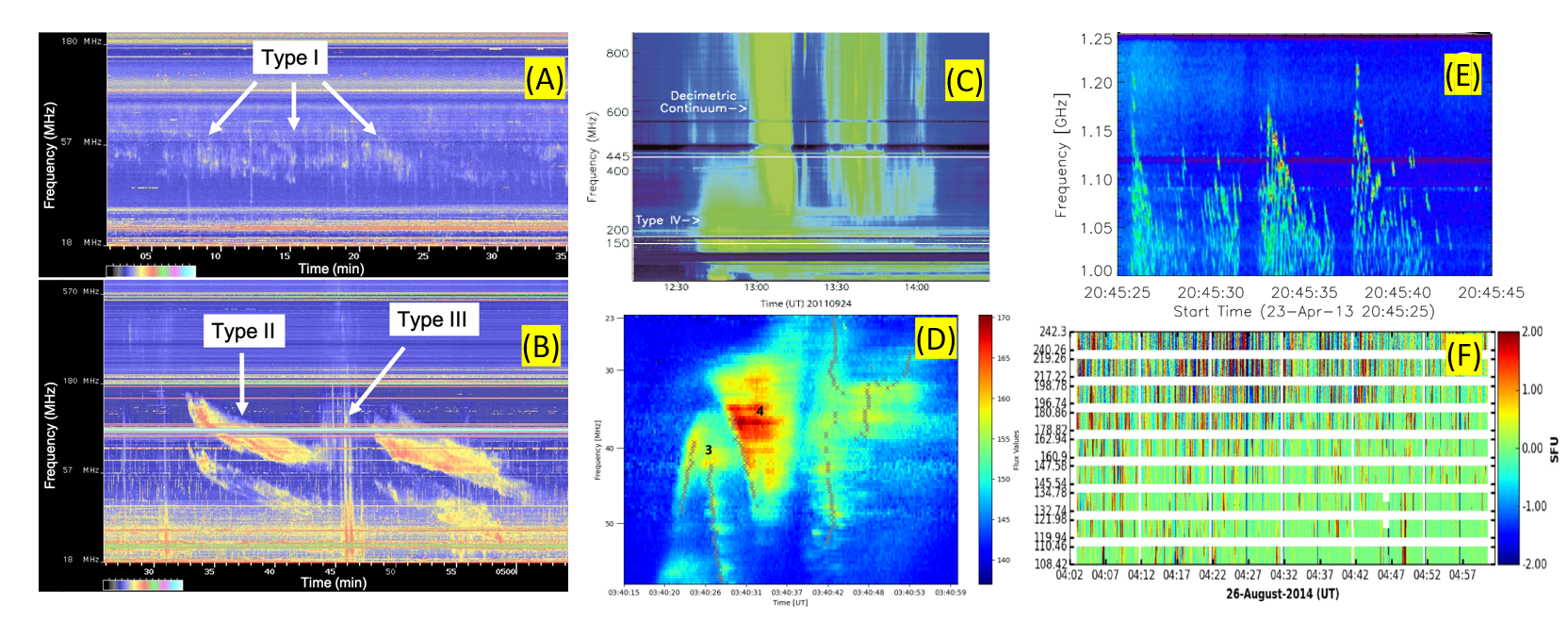} 
	\end{tabular}
    \caption{Traditional classification in the frequency-time plane of solar radio bursts from type-I to type-V, along with other types of bursts. Panel A: Type-I burst characterised by fragmented emission lasting for many hours. Panel B: Type-II and Type-III bursts showing slow and fast drifts in frequency. Panel C: Type-IV burst along with the wideband continuum. Panel D: Type-V burst marked with letter 4, while the U-burst with distinct frequency reversal is marked with number 3. Panel E: The dynamic spectrum of microwave spike bursts temporally occurring in bunches. Note the fine structures within the radio bursts. Panel F: Running mean subtracted flux calibrated dynamic spectra showing ubiquitous low-level (< 1 SFU) impulsive radio bursts present at low frequencies from MWA. Note that 1 SFU = 10$^4$ Jy. The dynamic spectra are taken from \cite[][]{2019arXiv191201747M,2024SoPh..299..146B,2018ApJ...852...69S,2021ApJ...922..134B,2018SoPh..293...58L}}
    \label{fig:bursts}
\end{figure*}

Radio wavelengths are excellent probes of the nonthermal population \citep{1980gbs..bookR....M}. Radio bursts are intense flashes of radio waves, usually compact in spatial extent \citep{1985srph.book.....M}. Before the advent of radio interferometers, the classification of radio waves historically occurred based on their size and shapes in the frequency-time plane. There are 5 main types of radio bursts, each one distinct in the frequency-time plane, and originating from different types of processes \citep{1965sra..book.....K}. Fig. \ref{fig:bursts} shows the frequency-time variations of radio bursts. Type-I bursts track the continuous acceleration and particle trapping in larger coronal loops. Type-II bursts are produced by particles accelerated during the shock propagation, while type-III bursts originate from propagating electron beams in the solar corona, creating frequency drifts in the frequency-time plane \citep{2014RAA....14..773R} depending on their characteristic velocities. Type-III burst drifts ($\sim$ 10 sec) much faster than type-II bursts. They are used for estimating emission heights and tracing magnetic fields along the path of particle propagation \citep{2013ApJ...763L..21C}. Sometimes, the nonthermal electrons are observed to escape the solar atmosphere into the interplanetary medium \citep{2019A&A...632A.108M}. This leads to type-III bursts observed at very low frequencies ($<$ 1 MHz), connecting solar plasma diagnostics to space weather. Polarisation of type-III is an important measurement for deciphering emission mechanism, geometry or single vs multiple acceleration events\citep{2022SoPh..297...47M}. Type-IV bursts are believed to be from particle trapping in magnetic loops. The Type-V bursts are from the propagation of accelerated nonthermal electrons in the magnetic loops \citep{2024ApJ...971...86M,1980A&A....88..218D,2017A&A...606A.141R}. In addition, the polarization measurements of type-IV bursts helps in understanding the evolution of the sources. \citep{2020A&A...639A.102S}



\begin{figure*}[h]
    \centering
    \begin{tabular}{c}
	\includegraphics[width=0.7\columnwidth]{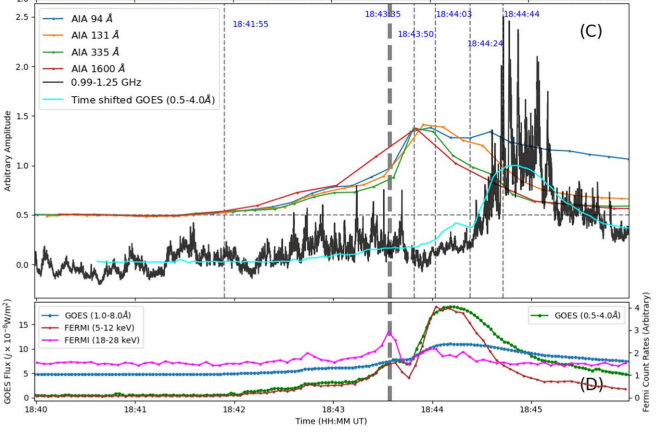} 
	\end{tabular}
    \caption{Lightcurves derived for a solar flare. Top panel shows AIA lightcurves obtained from integrating the flare region. The black curve shows the lightcurve for the radio bursts. The bottom plot shows the Hard X-rays in magenta, while soft X-rays from the FERMI lightcurve are shown in brown. Note that the hard X-ray emission peaks first, while the soft X-ray and radio lightcurves shows a significant time delay of $\approx$28 sec. More details here: \cite{2024ApJ...970...17S}}
    \label{fig:lightcurve}
\end{figure*}

\subsection{Spikes \& other types of radio bursts}

In the last few decades, high-resolution spectroscopy, due to improved digital receivers, has revealed previously unobserved spectral-temporal variations with fine structures associated with radio bursts. With fine spectral resolution, another type of bursts known as `microwave spikes' can be captured \citep{1982SSRv...32..405K}. These are intense narrowband bursts with sub-second structures (Fig. \ref{fig:bursts}(D)) \citep{2021ApJ...922..134B}. The fine structures are present at milli-second timescale and are high circularly polarised \citep{1992A&AS...93..539B}. These are proposed to be from multiple emission mechanisms like electron-cyclotron maser, plasma emission etc.. \citep{1978Natur.275..520S, 2024ApJ...969....3W}. Spikes at meterwaves may be associated with acceleration of electrons, i.e. also in the vicinity of type-III bursts \citep{2023ApJ...946...33C}. With recent developments in interferometry with dense cores of antennas, ubiquitous weaker impulsive radio emission has been observed, adding to the puzzle of the particle acceleration (Fig. \ref{fig:bursts}(F)) \citep{2022ApJ...937...99S}. Spikes are one such radio burst which occur in a wide frequency range covered by all frequency bands of SKAO. Solar radio spikes are observed in association with solar flares and coronal mass ejections (CMEs), and only recently could be imaged resolving in frequency and time \citep{2021ApJ...917L..32C,2023ApJ...946...33C}.

Spikes show clustering of many smaller emission structures ($>10^4$) spread over minute timescales. They appear to be distributed stochastically \citep{1994SSRv...68..185I}. The clusters have intricate structures in frequency, e.g. harmonics and drifts in time, along with fast time variations at milliseconds \citep{1986SoPh..104...99B}. In terms of location, only some spike sources have been investigated due to the unavailability of simultaneous wide frequency imaging for the Sun or the limited frequency range of existing instruments \citep{2002A&A...383..678B}. The spatial origin of spike sources is a puzzle, as many of the spike locations do not correlate with the high-energy X-ray source in the conventional flare model \citep{2009A&A...499L..33B, 2021ApJ...922..134B}. However, total flux density and burst rate of spike clusters do correlate with hard X-rays with a time delay \citep{1992ApJ...401..736A}. Further, the fragmentation of energy in spikes and their emission mechanism is a puzzle \citep{1985SoPh...96..357B}. 

Some studies suggest plasma emission and ECM from a narrow bandwidth of the spikes at the local gyrofrequency of the electron and its harmonics \citep{ABALDE2001741}. The first imaging-spectroscopy studies \citep[e.g.][]{2021ApJ...917L..32C} suggest plasma emission of the spikes. Scattering of radio wave propagation is another factor which impacts measured timescales of spikes \citep{2017NatCo...8.1515K}. Correcting for scattering \citep{2023ApJ...956..112K}, the actual energy release timescales may be much shorter (tens of milliseconds) than previously assumed due to the dominance of scattering in shaping observed profiles \citep{2023ApJ...946...33C}. The high-time-frequency resolution imaging spectroscopy with the SKAO will extend these studies to higher frequencies (up to several GHz) and with much higher sensitivity, which will provide crucial information about the plasma conditions and emission mechanisms in the solar corona.

\subsection{Ubiquitous non-thermal emission}

The classification of the intense solar radio bursts and their association with solar flares implied a natural distinction of `quiet' and `active' time phases \citep{RAULIN2005739}. However, in the last decade, the notion of `quietness' of the Sun has been challenged with more sensitive and high time resolution data e.g. MWA \citep{2017SoPh..292...75O,2018ApJ...852...69S,2023ApJ...943..122M}. There is evidence for a weaker level of solar activity ubiquitously present in time and located all over the solar disk \citep{2022ApJ...937...99S,2020ApJ...895L..39M}. Fig. \ref{fig:winsque} (B) shows the location of weak radio impulsive emission spread all over the solar disk, with some regions more concentrated. The frequency-time spans of the weak impulsive features reveal unresolved structures longer than 4 MHz bandwidth and 0.5 sec time resolution (Fig. \ref{fig:winsque} (C)). More sensitive and high spatial resolution data along with polarisation are expected to facilitate the localisation of such weak features and aid in spatio-temporal comparisons with EUV.  

\subsection{Gyrosynchrotron Emission}

High-energy electrons with tens to hundreds of keV gyrating magnetic fields give rise to GS emission in microwaves. They originate in association with solar flares, coronal mass ejections, etc \citep{2016ApJ...826...38F}. These gyrating electrons, travelling in the flaring loops, emit GS emission in radio frequencies. During the solar flare, they can reach $>$100 keV energies, which dominate the radiation spectrum (above a few GHz) \citep{1986SoPh..104..207A}.  These GS sources exhibit broadband emission ranging from a few 100 MHz to hundreds of GHz, i.e., relevant for SKAO observations. The characteristic GS spectrum has optically thick and thin frequency ranges. The GS fits various plasma parameters like magnetic field strengths, electron density, power-law index, etc. Fig. \ref{fig:gs} (B) shows a typical gyro-synchrotron during the solar flare, showing the peak at around 3 GHz. Optically thick and thin are below and above the peak, respectively. The constraints on magnetic field, nonthermal density and electron power-law, etc.. are mentioned. These parameters influence the shape and peak of the GS spectrum, e.g. the frequency location of the GS fits is determined by the magnetic field strengths. The electrons producing the GS emission can be either a non-thermal or a thermal distribution heated by a solar flare. GS emission can be produced in multiple regions of the magnetic loops, e.g. in the acceleration region or away from it due to propagation or via trapped electrons and accumulation at the loop top \citep{1989ApJ...339.1115G, 2011ApJ...731L..19F}.

\subsection{Open questions in Solar flare physics relevant to radio wavelengths}

The lack of understanding of particle acceleration results in questions surrounding a multitude of solar phenomena. However, some pointed questions arising from the direct involvement of particle accelerations are the following. 

\begin{enumerate}
\item Detailed model of electron transport: Despite having a basic particle transport model, there are open questions concerning energetic electron acceleration, trapping, release in the corona, and transport in interplanetary space. 

\item Origin of radio spikes sources: The observed source size of solar radio spikes is compact; however, the absolute position with respect to the active region geometry is ill-defined. 

\item Coronal Heating in higher Corona: How does energy release in the higher corona impact the coronal heating?
\item How frequently particle acceleration occurs in the quiescent Sun, and where does weak particle acceleration take place?
\end{enumerate}

SKAO observations of the Sun will help answer many aspects of these questions, as discussed in the following sections.

\begin{figure*}[h]
    \centering
    \begin{tabular}{ccc}
	\includegraphics[width=0.3\columnwidth]{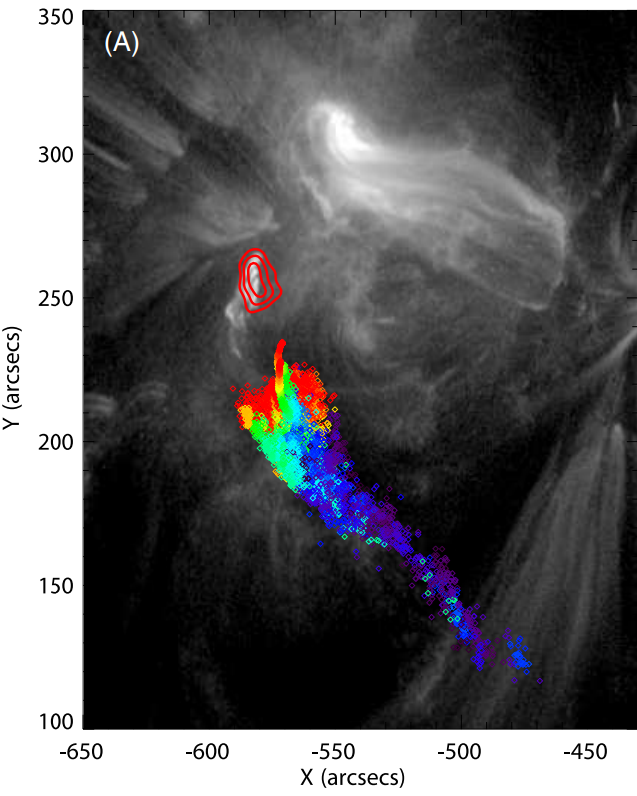} &
	\includegraphics[width=0.25\columnwidth]{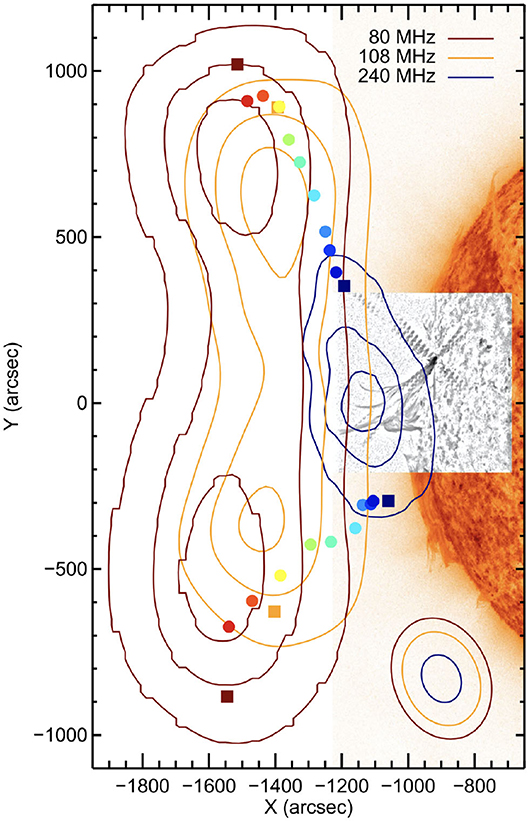} &
	\includegraphics[width=0.4\columnwidth]{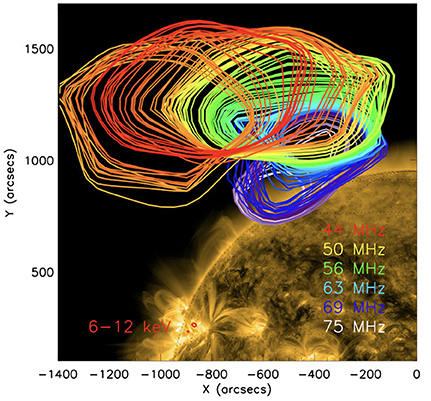} \\
	(A) VLA type-III burst & (B) MWA Contours & (C) LOFAR Contours 
	\end{tabular}
    \caption{Science case C1 (see Sec. \ref{sec:ska}): The spectroscopic imaging of the type-III and U-radio bursts tracing the magnetic field topology in the solar corona. Panel A: Centroid location of the radio sources for different frequencies tracing local magnetic field \citep{2013ApJ...763L..21C}. The red contour is the X-ray source. Panel B: The radio source split at low frequencies is attributed to the magnetic field geometry \citep{2017ApJ...851..151M}. Panel C: The radio contours from a U-burst tracing the magnetic loop top \citep{2017A&A...606A.141R}. }
    \label{fig:typeiii}
\end{figure*}

\begin{figure*}[h]
    \centering
    \begin{tabular}{ccc}
    	\includegraphics[width=0.3\columnwidth]{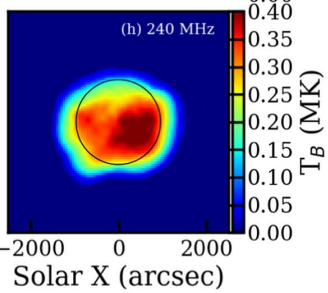} &
    	\includegraphics[width=0.35\columnwidth]{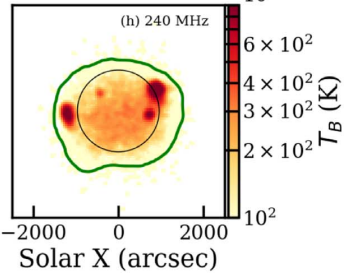} &
    	\includegraphics[width=0.35\columnwidth]{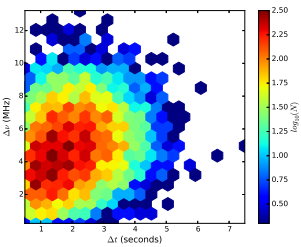}
    \end{tabular}
    \caption{Science case C2 (see Sec. \ref{sec:ska}): Characteristics of the weak impulsive emission from the Sun. Panel A: The radio image of the Sun at 240 MHz from MWA. Panel B: Corresponding to the image in the panel A, a time-averaged brightness temperature distribution of the weak impulsive emission. This is obtained by removing the static component from the full disk solar images. Note the low-brightness temperature levels and some concentration of the impulsive emission. Panel C: Frequency-time spans of the weak impulsive emissions. The colorbar indicates the counts of the features. We note that most features have a bandwidth of $\approx$4 MHz and a time span of $\approx1$ sec. Adapted from \cite{2017ApJ...843...19S,2022ApJ...937...99S}}
    \label{fig:winsque}
\end{figure*}

\section{Diagnostics of Particle Acceleration and Transport}
\label{sec:diagnostics}

Radio bursts and Hard X-rays are direct signatures of nonthermal particle populations. They indicate efficient particle acceleration processes in solar flares \citep{2023ApJ...947L..13K}; however, the spatial variation of the electron spectrum confirms a small fraction of electrons are involved in acceleration \citep{2025ApJ...990..101V}. 
In the standard flare model, the particles are accelerated near the magnetic reconnection sites and then travel to faraway locations on the solar disk guided by the surrounding magnetic fields \citep{2013ApJ...771...82M}. Using radio emission, we can probe regions close to the reconnection site, as well as the transport of the electron beams along magnetic fields \citep{2020ApJ...904...94S,2024NatAs...8...50Y}. The propagation of electron beams can be traced by type III, type U, and reverse-drifting type III radio bursts \citep{2013ApJ...763L..21C,2025NatSR..1511335C}. The latter shows evidence of oppositely propagating electron beams in upward and downward directions, creating a pair of radio bursts \citep{2000SoPh..194..345R}. 

Solar observations at SKA-low frequencies provide a unique probe of the high solar corona, i.e. the region of particle escape into the interplanetary medium via solar wind and CME shocks. CME shocks are initiated during solar flares. Moving type-IV bursts indicate signatures of trapped particles emitting gyrosynchrotron emission \citep{2017A&A...608A.137C}. At SKA-mid radio frequencies, the acceleration region, along with the acceleration mechanism, can be probed \citep{1984SoPh...90..383D, 1998JGR...10317223D}. They provide constraints for the processes of magnetic reconnection, including CME shock initiation. However, the precise details of these scenarios are yet to be understood.


\subsection{Particle acceleration}

Solar energetic particles study offers an opportunity to understand high-energy acceleration processes. Particle acceleration can be studied by measuring the plasma parameters—such as the electron and ion density, temperature, velocity, and the magnetic field—during a solar flare \citep{2026ApJ...999..179F}. The locations of the particle acceleration, where the magnetic reconnection takes place to drive the solar flare, remain a mystery. GS emission estimates them near the energy release sites.  In coronal mass ejection, there is a lot of variation in the magnetic field across its geometry and launch, i.e., a wide contrast of magnetic fields has been observed \citep{2024ApJ...968...55K}. In addition, from microwave observations, the evolution of the spatially resolved distributions of thermal and non-thermal electrons in a solar flare has demonstrated highly efficient electron acceleration in the acceleration region \cite{2022Natur.606..674F}.
In conjunction with X-rays, microwave spectroscopic imaging allows us to probe multiple electron populations from solar flares or CMEs \citep{2020ApJ...895L..50C,2021ApJ...908L..55C}.
\subsubsection{GS Spectrum}
Fig. \ref{fig:gs} (D) shows a variety of GS spectra spread throughout the SKAO frequency ranges, indicating a rich diversity of magnetic field structures. With SKAO observation, the measurements across this wide frequency range will be robust and comprehensive. Further, the spatial evolution along with GS fits is a powerful tool for understanding the dynamics of particle acceleration. Therefore, the spatial and spectral domains need to be well-sampled.




\subsection{Re-acceleration}
Solar flare accelerated electrons propagating along field lines could not only loose energy but gain energy. So the electrons after their initial acceleration, are accelerated again \citep{2009A&A...508..993B}. This can happen in various parts of the solar flare, such as in turbulent regions or current sheets, and can be caused by phenomena like Langmuir waves or other stochastic processes. Re-acceleration of propagating electron beams due to collective effects is closely related to the evolution of the Langmuir wave spectrum towards smaller wavenumbers \citep{2012A&A...539A..43K,2012A&A...544A.148K}.
Hence the plasma radio emission from flaring loops is expected \citep{2014A&A...562A..57R}.
The numerical simulations \citep{2012A&A...539A..43K} emphasize that collective plasma effects should not be treated simply as an additional energy-loss mechanism, when hard X-ray emission in solar flares is interpreted. The redistribution of beam energy in propagating electron beams results in a local heating maximum that is similar to a radiative-hydrodynamic model with a large, low-energy cutoff and a hard power-law index and explain the color temperatures and Balmer jump strengths in stellar flare observations\citep{2023ApJ...943L..23K}.
Importantly, the Langmuir waves and characteristic plasma emission 
can be observed to test the efficiency and the details of electron acceleration models. Therefore, using the high resolution spectrometry would be useful in constraining the parameters of beam energy using radio emission. SKAO's wide frequency and high spatio-spectral and temporal resolution will help in constraining the re-acceleration special cases. 

\subsection{Particle transport}

Understanding particle propagation from the reconnection site to the radiation regions through observations is challenging. In the standard flare model, the timing of the hard X-ray observations from the base of loops will have a propagation delay \citep{2024MNRAS.532..705S}. Similarly, the radio emission will also have a propagation delay \citep[e.g.][]{2024FrASS..1107955K}. The delay depends on the magnetic topology, energy of the particles, and emission mechanisms \citep{2024ApJ...970...17S}. Fig. \ref{fig:lightcurve} shows an example of radio bursts and X-ray emission showing a time delay, which allows us to constrain the electron propagation paths assisted by magnetic extrapolation models. Even though radio and X-ray producing electrons have different energies, due to the time correlation, it is assumed that they are part of the same accelerated electron population. The strong evidence of close hard X-rays and radio bursts timings suggests a precipitating electron beams interpretation of particle transport. Various processes like magnetic mirroring, convoluted magnetic topology, pitch angles and energy distribution create ambiguity in determining simple delays, i.e. well-sampled spatial-temporal observations are required.


\begin{figure*}[h]
    \centering
    \begin{tabular}{cc}
	\includegraphics[width=0.4\columnwidth]{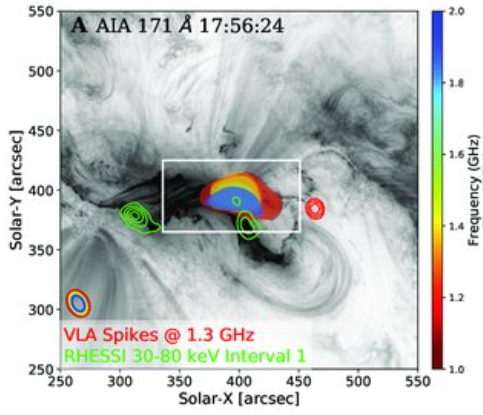}&
	\includegraphics[width=0.4\columnwidth]{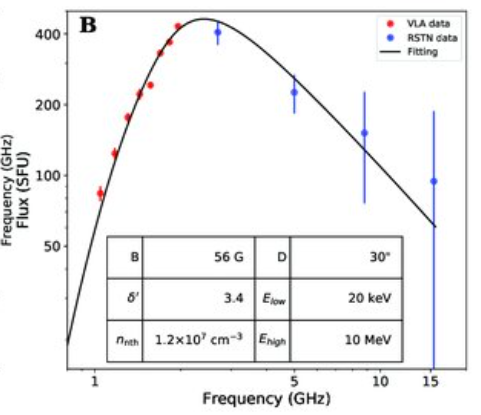}\\
	(A) VLA Contours & (B) Flare's GS spectrum \\
	\includegraphics[width=0.4\columnwidth]{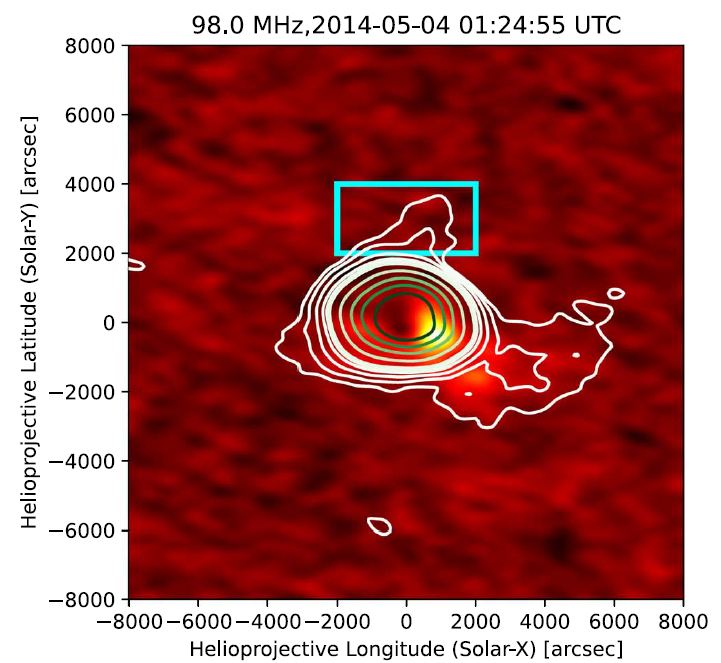}&
	\includegraphics[width=0.45\columnwidth]{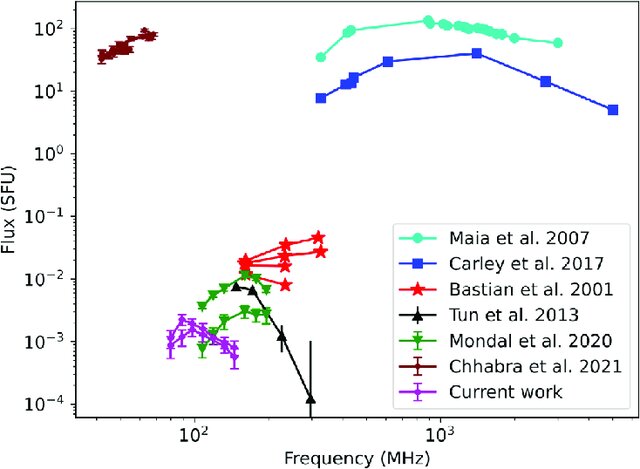}\\
	(C) MWA Contours & (D) CME's GS spectrum \\
	\end{tabular}    
\caption{Science case C3-C4 (see Sec. \ref{sec:ska}): Example of spatial-spectroscopic observation showing the GS spectrum in a solar flare observed by VLA, and GS fits for CME observed by MWA Panel A: X-ray (green) and radio (rgb solid) contours overlay on the EUV 171 $\AA$. Panel B: The GS frequency spectrum partly captured by VLA and rest filled-in by RSTN non-imaging measurements.  Panel C: Background image is the Stokes V image, while white contours are Stokes I for CME at 98 MHz. Note that zero Stokes V emission is detected from the CME in the cyan box. Panel D: Comparison of spectra of GS emission from CME plasma. The magenta points represent sample spectra from regions in panel C. Note the wide frequency-range of GS emission. Adapted from \cite{2021ApJ...911....4L,2023ApJ...950..164K}}
    \label{fig:gs}
\end{figure*}

\section{Improvements with SKAO}

\label{sec:ska}





\subsection{Status from SKAO's Precursor and Pathfinder telescopes}

The radio interferometers have evolved over decades, improving the large field-of-view and sensitivity at low frequencies. Recent interferometers like LOFAR \& MWA explore the so-called large-N and small-D array concept that is also used for the SKAO \citep[e.g.][]{2009IEEEP..97.1497L}. One of the requirements for high-fidelity solar imaging is a dense network of smaller baselines. However, at high frequencies ($\geq$ 500 MHz), the configuration with a dense network is particularly hard due to dish-based array design and cost. Low-frequency SKAO's precursor, like MWA and LOFAR, has led to significant improvement in the spectroscopic imaging. Significant strides have been made in polarisation-related studies and measurement using MWA \citep{2022ApJ...932..110K,2025ApJS..278...26K}. Solar observation from Meerkat precursor has produced high-quality solar imaging \citep{2025FrASS..1266743K,2026ApJ...998L..46L}. Some of the major challenges in solar imaging spectroscopy that remain are the following.
\begin{itemize}
\item In most radio events, the spectroscopic data is unable to capture the whole event. Either the events are clipped in frequency spans, spatially or temporally unresolved. 
\item Solar flare and radio bursts are unpredictable, i.e. they mostly get missed from non-solar dedicated radio interferometers
\item Capturing fine structures in frequency-time spans of the radio bursts and spikes is limited by the instrumentation
\item Good calibrated polarization measurements for radio bursts are difficult
\end{itemize}

\begin{table}
\centering
\begin{tabular}{|c|c|c|c|c|}
\hline
SKAO band & Start (MHz) & End (MHz) &  Primary Beam \\ 
\hline
Low & 50 & 350 &  1.23$^{o}$ \\
Band-1 & 350 & 1050  & 1.1$^{o}$ \\
Band-2 & 950 & 1760  & 0.65$^{o}$ \\
Band-3 & 1650 & 3050  & 0.38$^{o}$ \\
Band-4 & 2800 & 5180 & 0.22$^{o}$ (792") \\
Band-5a & 4600 & 8500  & 0.13$^{o}$ (468")\\
Band-5b & 8300 & 15300  & 0.075$^o$ (270")\\
\hline
\end{tabular}
\caption{Frequencies for different bands of SKAO. The primary beam scales on the sky computed for highest frequency for different SKAO band.}
\end{table}

\begin{table}
\begin{tabular}{|c|c|c|c|c|c|}
\hline
Parameters & VLA & EOVSA & MWA & LOFAR & NRH \\
\hline
Frequency Span (GHz) & 1-4 & 1-18 & 0.08-0.3 & 0.02-0.08 & 0.15-0.45\\
Typical Spatial Resolution & 19" & 3-57" & 180"  & 60"-540" & 18-240" \\
Time Resolution (ms) & 50 & 20 & 500 & 10 &250 \\
Number of Antennas & 27 & 15 & 128 & 52 & 44\\
Polarimetry & Yes & Yes & Yes & Yes & Yes\\
\hline
\end{tabular}
\caption{Specifications of VLA imaging instruments used for particle acceleration studies. Adapted from \cite{2019AdSpR..63.1404N}}.
\end{table}

These challenges require a well-characterised instrument with wide frequency observation coverage, with dense UV coverage observing the Sun continuously. SKAO precursors and pathfinders have addressed some of these challenges with their own limitations. For e.g., at SKA-low frequencies, non-solar dedicated MWA and LOFAR have shown high fidelity images of the solar disk with limited angular resolution. At SKA mid-frequencies, the expanded Owens Valley Solar Array (EOVSA, \cite{2012AAS...22020430G}) and Very Large Array (VLA, \cite{2009IEEEP..97.1448P}) have provided excellent resolution and time resolution for the radio bursts, with limited UV coverage.
SKAO can address many of the challenges in pursuit of ideal solar observations. 

\begin{table}[h]
	\centering
	\scriptsize
\begin{tabular}{|l|cccccc|}
\hline
Representative freqeuncy & 110 MHz & 300 MHz & 770 MHz & 1.4 GHz & 6.7 GHz & 12.5 GHz \\
\hline
Range [GHz] & 0.05--0.35 & 0.05--0.35 & 0.35--1.05 & 0.95--1.76 & 4.6--8.5 & 8.3--15.3 \\
\hline
Telescope & Low & Low & Mid & Mid & Mid & Mid \\
\hline
FoV [arcmin] & 327 & 120 & 109 & 60 & 12.5 & 6.7 \\
\hline
Max. Resolution (arcsec) & 11 & 4 & 0.7 & 0.4 & 0.08 & 0.04 \\
\hline
Max. Bandwidth [MHz] & 300 & 300 & 700 & 810 & 3900 & $2\times 2500$ \\
\hline
Channel width [kHz] & 5.4 & 5.4 & 13.4 & 13.4 & 53.8 & 80.6 \\
\hline
Spectral zoom windows $\times$ narrowest BW [MHz]$^d$ & $64\times 0.025$ & $64\times 0.025$ & $4\times 3.1$ & $4\times 3.1$ & $4\times 3.1$ & $4\times 3.1$ \\
\hline
Finest zoom channel width [Hz] & 14 & 14 & 210 & 210 & 210 & 210 \\
\hline
\end{tabular}
	\caption{SKAO's Expected Performance – for design baseline of AA4. Note that the noise in the images will be limited by self-noise for the solar observations \citep{1989AJ.....98.1112K}. }
\label{tab:specifications}
\end{table}

\textit{SKAO's Improvements:}

SKAO will provide unprecedented data quality both at SKA-low and SKA-mid frequencies. SKA-Mid will consist of 133 SKAO dishes with 15 m diameter and 64 Meerkat dishes with 13.5 m diameter. Table \ref{tab:specifications} shows the specifications for the SKAO at SKA-low and SKA-mid frequency bands. For a 15 m SKAO dish, a typical solar radio disk of $\sim$0.6 deg angular scale, the frequencies $>1.9$ GHz will resolve the solar disk. For SKA-mid science cases, for full solar disk sub-array beam modes, and mosaic stitching will be needed. The SKA-mid solar science cases will be designed to be focused on the regions of the Sun, like active regions, limb regions, etc... For SKA-low, the station size is $\sim$40 m, and the primary beam will capture the entire solar disk. Given the SKAO configurations, some of the key improvements in future SKAO solar images are the following. 
\begin{enumerate}
\item Wide frequency capture of the radio emission
\item High-resolution imaging spectroscopy 
\item Improved sensitivity/fidelity 
\item Full polarisation imaging
\end{enumerate}

In the range of SKAO radio frequencies (50 MHz to 15 GHz) captures different accelerated particle populations are captured in different bands. An extreme example could be near-simultaneous type-III bursts and radio spikes in an eruptive event may occur in SKA-low and SKA-mid frequencies separated by 100s of MHz. Hence, a wide frequency coverage is an important advantage to capture different particle acceleration events from the same solar flare, jet eruption, etc.. High spatial resolution images and spectroscopic images from SKAO with full Stokes polarisation will improve the location constraints of the radio burst and plasma diagnostics from precursor telescopes ( Table \ref{tab:baseline}). Further, dense UV coverage is desired to capture source morphology, especially on the solar disks with a variety of features. As we discussed in the earlier section \ref{sec:diagnostics}, there are aspects of particle acceleration that can be studied using radio wavelengths. For this paper, we chose the four most interesting cases (C1 to C4). 


\begin{enumerate}

\item C1: Tracing magnetic fields with type-III bursts

\item C2: Localisation of the weak impulsive sources w.r.t EUV counterparts 

\item C3: Constraining magnetic fields and other plasma properties during a solar flare

\item C4: Diagnosis of magnetic field evolution during a CME launch

\end{enumerate}

These areas capture the majority of the particle acceleration studies (see in Fig. \ref{fig:typeiii}, \ref{fig:winsque} \& \ref{fig:gs}). However, there are other particle acceleration problems which will also benefit from the improvement of the SKAO mentioned above. 

\subsection{Progressive improvements in Array Assemblies (AA)}

\begin{table}[htbp]
\centering
\tiny
\begin{tabular}{|l|cc|cc|cc|cc|cc|cc|}
\hline
\textbf{Parameters} & \multicolumn{2}{c|}{\textbf{AA0.5}} & \multicolumn{2}{c|}{\textbf{AA1}} & \multicolumn{2}{c|}{\textbf{AA2}} & \multicolumn{2}{c|}{\textbf{AA*}} & \multicolumn{2}{c|}{\textbf{AA4}} & \multicolumn{2}{c|}{\textbf{Precursor}} \\
\cline{2-13}
& \textbf{Low} & \textbf{Mid} & \textbf{Low} & \textbf{Mid} & \textbf{Low} & \textbf{Mid} & \textbf{Low} & \textbf{Mid} & \textbf{Low} & \textbf{Mid} & \textbf{MWA} & \textbf{Meerkat} \\
\hline
Number of antennas & 6 & 6 & 120 & 28 & 2016 & 2016 & 46971 & 10296 & 130816 & 19306 & 8128& 2016 \\
\hline
 Maximum Baseline (m)& 5716 & 1415 & 44883 & 1415 & 61288 & 106550 & 65369 & 106550 & 65413 & 157397 & 2874 & 7698 \\
\hline
Band-low (") & 54.13 & - & 6.89 & - & 5.05 & - & 4.73 & - & 4.73 & - & 107.67 & - \\
Band-mid1 (") & - & 62.45 & - & 62.45 & - & 0.83 & - & 0.83 & - & 0.56 & - & 11.48 \\
Band-mid5a (") & - & 6.67 & - & 6.67 & - & 0.09 & - & 0.09 & - & 0.06 & - & 1.23 \\
\hline
Median Baseline (m) & 4246 & 1102 & 3683 & 665 & 3324 & 2124 & 1229 & 2223 & 4904 & 2838 & 622 & 957 \\\hline
Band-low (") & 72.87 & - & 84.01 & - & 93.08 & - & 251.79 & - & 63.09 & - & 497.59 & - \\
Band-mid1 (") & - & 80.19 & - & 132.98 & - & 41.62 & - & 39.77 & - & 31.14 & - & 92.32 \\
Band-mid5a (") & - & 8.57 & - & 14.21 & - & 4.45 & - & 4.25 & - & 3.33 & - & 9.87 \\
\hline
\end{tabular}
\\

\caption{Configuration information for various array assemblies of SKAO and precursors. The maximum baseline, median baseline and corresponding angular scales (in arcsec) for both SKA-low and SKA-mid array assemblies. The baseline lengths and corresponding angular scales on the sky. The angular scale is computed for the solar angular scale of 0.6 degrees on the sky, and at the mid-band frequency.}
\label{tab:baseline}
\end{table}

The SKAO's array assemblies (AA) progressively improve in the number of stations or dishes, bandwidth, frequency, and time resolutions etc... For SKA-low, the large baselines are under construction first, followed by the smaller baselines. This will be reflected in a gradual increase in the fidelity of the solar images. Table \ref{tab:baseline} shows the improvement in the number of baselines, resolution, and median baseline length and resolution for various array assemblies.

At SKA-low frequencies, for the initial array assemblies (AA0.5 to AA2), the absence of the smaller baselines will make the imaging of the extended solar disk difficult. E.g., case 2 requires dense baselines, helpful in detecting the fine flux density variations over the solar disk. The SKAO array assembly AA* \& AA4 will have several baselines greater than MWA. In addition, the median baseline lengths of array assemblies are higher than MWA, leading to better resolution. On the other hand, the compact radio bursts and active regions will be the focus during the initial phase (AA0.5 to AA2). With AA* and AA4, with more dense UV-coverage, we can expect high dynamic range imaging, as MWA achieved images of $\approx 10^6$ dynamic range. In MWA images, the high dynamic range enabled observing the quiet Sun in the presence of a burst. This is particularly important for simultaneous studies of the multiple active regions present on the disk. At SKA-low frequencies, the radio bursts related studies, like magnetic field tracing and gyrosynchrotron plasma diagnostics, rely on good sampling in the frequency domain along with high fidelity imaging. A good description of the morphology of radio bursts enables spatially resolved plasma diagnostics, i.e., mapping of the magnetic fields, electron density, etc... However, at SKAO, low-frequency scattering and refraction can broaden and shift the source locations, thereby probing underlying coronal turbulence. As the number of baselines increases considerably from AA1 to AA2, i.e., AA2 onwards, the morphology is expected to improve, i.e., from AA2 to AA4, enabling case-1, case-3, and case-4.

For SKA-mid, the solar observations can be viewed in the different bands. The SKA-mid dish sizes will resolve the solar disk above $\approx$2 GHz. Therefore, sub-array modes operational mode is required from band-2 onwards. Band 2 will barely resolve the solar disk. To cover the solar disk, the sub-arrays need to be pointed at different regions on the disk. As there are 16 sub-arrays possible, the regions of interest need to be defined for possible capture of the radio bursts. For C3, the spatially resolved gyrosynchrotron spectrum can be obtained in unprecedented spatial detail. From the AA2 configuration, the spatial resolution will reach arcsec and sub-arcsec angular scales, comparable to the EUV image resolution. By AA4 deployment, the number of baselines also picks up, and good morphology will help trace coronal features within active regions. This advantage will enable accurate mapping from multi-wavelength studies of the energy release sites of the radio bursts or particle acceleration. 

\begin{figure*}[h]
    \centering
    \begin{tabular}{c}
	\includegraphics[width=1.0\columnwidth]{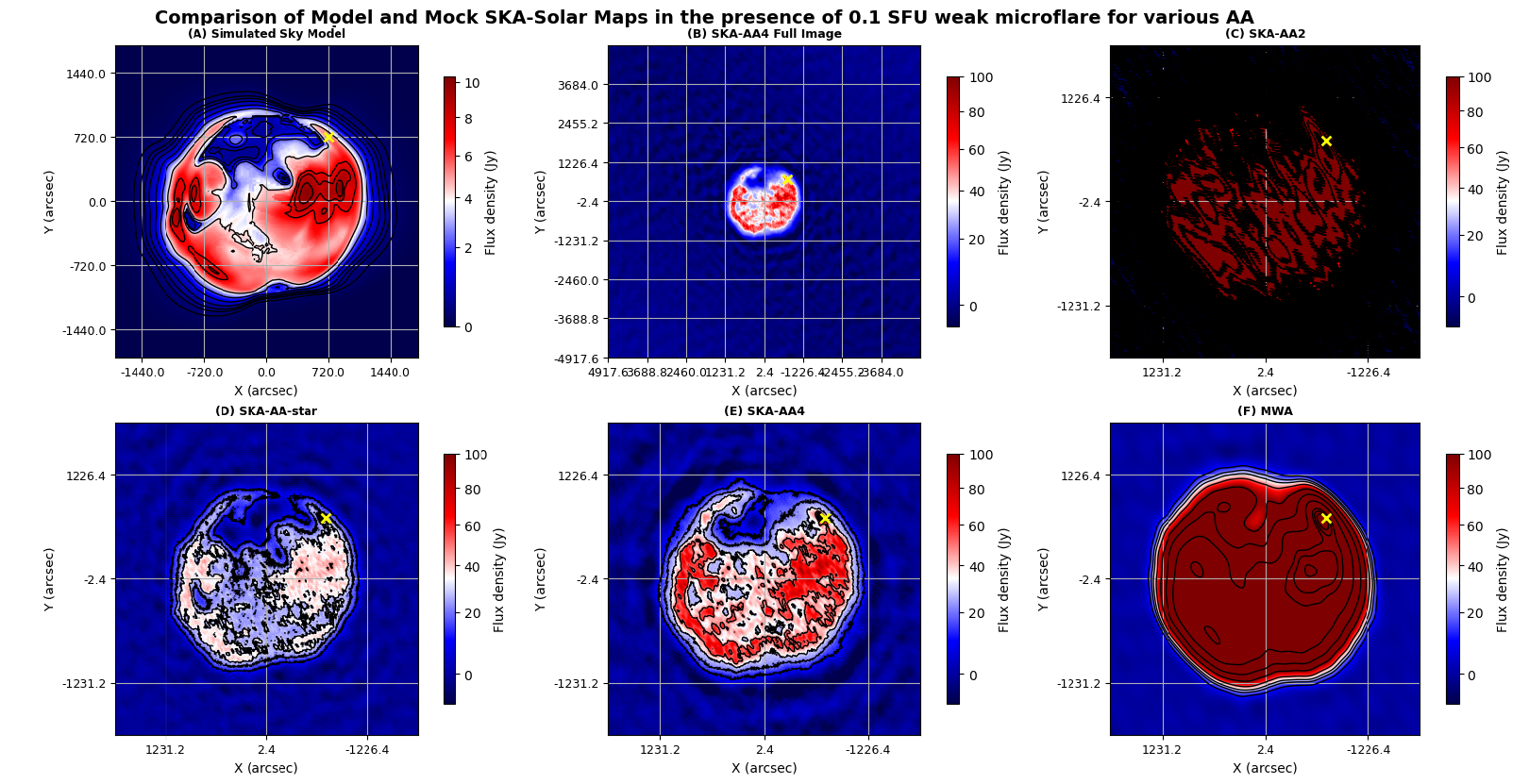}
	\end{tabular}    
\caption{Simulated image of a weak solar burst (0.1 SFU) in the presence of the solar disk emission. Panel A shows the FORWARD solar map in flux densities showing extended solar disk. The brightness temperatures were converted into flux densities, assuming a pixel-sized source emission region. The pixel size for the FORWARD map is 22.5". The black contours are at 0.9, 0.8, 0.7, 0.4, 0.2, 0.1, 0.05, 0.03, 0.02, 0.01 fraction of the maximum flux density in the map, and are the same for all panels. Note that the radio burst source is not shown in the colorbar in panel A. However, the location of the radio burst source is marked in a yellow cross in each panel. Panel B shows the full field-of-view reconstructed map for the AA4 configuration. Panels C, D, and E show the zoomed reconstructed map for AA2, AA*, and AA4, respectively. Panel F shows the reconstructed map for the MWA configuration. In all reconstructed maps, the colorbar is saturated to 100 Jy to highlight the solar disk features, while the radio burst source is present at 1000 Jy. Note the improvement in solar disk features from MWA \& AA2 to AA4.}
    \label{fig:sim}
\end{figure*}

\subsection{Simulation of a weak radio burst with SKAO-AA}

Radio imaging with bright, extended sources is challenging due to the complex brightness distribution in the images. High variability adds to the challenge, i.e., simulation tools provide a way to study the impact on radio observations. The main challenge in imaging the Sun, as discussed in previous sections, is to achieve high fidelity, i.e., the best description of the intrinsic brightness distribution in the image plane. SKAO will enable an accurate description of the solar features, i.e., both the faint solar disk and bright solar bursts. 

We do a simple test to capture the fidelity increment through SKAO array assemblies AA2, AA*, and AA4 configurations. We simulate a combination of a solar disk and a weak radio burst of flux density 0.1 SFU. Such weak radio bursts are weaker than traditional radio bursts but stronger than the extended disk emission. Therefore, the imaging of the stronger radio bursts will create artefacts on the solar disk; however, with improved SKAO AA configurations, the clarity of the disk features will become better. 

We use the \texttt{Karabo} simulator to make mock radio observations \citep{2026A&C....5401004S}. The main inputs required in \texttt{Karabo} are sky model, telescope, observation settings, and interferometer settings. It will yield a visibility dataset, which will be deconvolved into images using WSClean. To simulate the solar disk in Stokes I, we make use of the FORWARD package \citep{2015IAUS..305..245G} using the free-free emission module and PSIMAS 3-D coronal configuration. The solar disk was added with an artificial 0.1 SFU radio burst source as a single pixel in the skymodel. We do not consider the effects of scattering or refraction in the skymodel. We use SKA-low telescope configurations for the simulations, and do not consider the effect of thermal and other types of signal processing noises. We focus on simulating the impact of sampling on the solar disk with a compact radio burst source. We also perform a run for the MWA configuration to benchmark it with a precursor telescope. The mock observation was done using 1 frequency channel and 1 time stamp, of 10 kHz and 1 minute, mimicking a snapshot imaging mode. The skmodel was placed at the local zenith, and images were made with 2.4" arcsec pixel resolution and a uniform weighting scheme.  

In total, we performed 5 runs, 4 for SKAO AA and 1 for MWA at 100 MHz central frequency. Figure \ref{fig:sim} shows the input solar disk, along with reconstructed images. The images are saturated at 100 Jy to witness the features from the fainter extended solar disk. We note a significant improvement in the solar disk features from AA2 to AA4. Panel B shows the full images and the noise floor for the AA4 configuration, with a good dynamic range ($> 1000$).  We note that AA* and AA4 configurations recover solar disk features, as compared to AA2 and MWA. MWA images could not capture the high-resolution features of the input maps. In addition, the contrast in the flux distribution within the disk is captured much better in AA4 compared to MWA. This is expected as AA4 has a larger number of larger baselines. The bright radio burst location is captured by all telescope configurations (see yellow cross). However, we also notice distortion in the brightness distribution in AA* and AA4 compared to the FORWARD input images. These artefacts are expected due to the presence of bright radio burst sources, and need to be quantified and are a subject of future research. The self-noise contributes to the images and is the biggest contributor to the distortion in the fidelity for the Sun \citep{1989AJ.....98.1112K,2025SoPh..300...91B,2025SoPh..300...90B}. The self-noise needs to be studied in the context of SKAO images. From this exercise, we conclude that the AA4 solar SKA-low images have the potential to showcase high-fidelity images even in the presence of the radio bursts. Thus, enabling more sensitive spatial-temporal measurements during radio bursts or otherwise.

\section{Synergies with other solar radio telescopes and instruments}
\label{sec:synergies}

Solar emission occurring in a wide electromagnetic spectrum captures different aspects of the accelerated particles. Therefore, synergies of radio emission with other wavelengths are key for studying particle acceleration from solar flares and CMEs. In view of SKAO, multi-wavelength studies will lead to high-impact results. However, SKAO must have key features to enable multi-wavelength studies. Below, we list a few capabilities required in SKAO observations and the user science platform, i.e., 
\begin{enumerate}

\item solar event triggering, 
\item EUV quicklooks \& analysis, and
\item synergies with solar dedicated telescopes \& space-based observatories.
\end{enumerate}

Firstly, multi-wavelength studies of the solar events are critical for building a detailed understanding of the events. Especially, EUV images provide the morphological context to the solar flare, while X-rays provide the energies of the accelerated particles involved. Space-based observatories capturing EUV and X-ray events will be needed to decipher the events of interest and further study them. The DEM maps obtained provide the diagnostics of the thermal plasma, while hard X-rays are used in characterising the non-thermal population. Muse and Solar-C missions will provide excellent EUV spectroscopy. Secondly, the solar flare and CMEs are not predictable, and observing the Sun all the time from non-solar telescopes is not possible. Therefore, apart from solar dedicated times, we would require triggers from solar monitoring to capture solar flares and radio bursts. The triggers can come from one of the SKA-low stations. A demonstration of generating triggers for MWA from Yamagawa radio telescopes has been done, and is a part of solar observing mode in MWA \citep{2024AGUFMSH53A2954P}. Thirdly, any ground-based radio telescope cannot monitor solar activity over 24 hours, i.e., studies requiring continuous solar features over a period of days would require continuous monitoring. SKAO, with a spread in geographical location, will provide better coverage; however, it is limited in frequency spans. Therefore, synergies with other radio telescopes will be useful for such studies.

For the SKA-low frequencies, the electron beams up in the solar corona producing type-III bursts need to be studied with X-ray emission, as discussed in section 2. The regions of radio emission at the SKA-mid frequencies are lower in the corona, much closer to the EUV emission heights, thus providing the opportunity to simultaneous energy releases and constraining the plasma parameters.

\section{Summary}
\label{sec:summary}

We present an overview of the particle acceleration studies from solar flares with a focus on the under-construction SKAO. Several studies discussed will thrive on full polarisation, high-resolution imaging spectroscopic capabilities of SKAO, supported by dense instantaneous UV coverage and wideband available for the observations. We note that the solar radio bursts are direct indicators of particle acceleration and show high contrast in the flux densities and frequency-time spans. As large baselines are constructed first, the localisation of the radio source will be better from initial AA2 onwards for both low and mid. The progressive improvement in UV coverage will enable good fidelity imaging for full solar disk cases. The fidelity of the solar disk in the presence of radio bursts will improve until AA4. Multi-wavelength observations will be crucial in driving productive particle acceleration-related studies. Overall, the SKAO will enable an excellent instrument for constraining plasma parameters during particle acceleration events happening on the Sun.



\bibliographystyle{abbrvnat-maxbibnames4}
\bibliography{aaskaii_template} 






\end{document}